\documentclass[conference]{IEEEtran}

\usepackage{cite}
\usepackage{amsmath,amssymb,amsfonts}
\usepackage{algorithmic}
\usepackage{graphicx}
\usepackage{textcomp}
\usepackage{xcolor}
\usepackage{hyperref}

\usepackage{tikz}
\usetikzlibrary{decorations.pathmorphing}
\usetikzlibrary{positioning}
\usetikzlibrary{shapes,arrows,decorations.pathreplacing}
\usetikzlibrary{calc}

\def\BibTeX{{\rm B\kern-.05em{\sc i\kern-.025em b}\kern-.08em
    T\kern-.1667em\lower.7ex\hbox{E}\kern-.125emX}}
\begin{document}

\title{Towards a Staging Environment for the\\ Internet of Things}

\author{
\IEEEauthorblockN{Jossekin Beilharz\IEEEauthorrefmark{1}, Philipp Wiesner\IEEEauthorrefmark{3}, Arne Boockmeyer\IEEEauthorrefmark{1}, Florian Brokhausen\IEEEauthorrefmark{3},\\
Ilja Behnke\IEEEauthorrefmark{3}, Robert Schmid\IEEEauthorrefmark{1}, Lukas Pirl\IEEEauthorrefmark{1}, and Lauritz Thamsen\IEEEauthorrefmark{3}}

\IEEEauthorblockA{
\IEEEauthorrefmark{1}
\{firstname.lastname\}@hpi.de, Hasso Plattner Institute, University of Potsdam, Germany\\
}
\IEEEauthorblockA{
\IEEEauthorrefmark{3}
\{wiesner, florian.brokhausen, i.behnke, lauritz.thamsen\}@tu-berlin.de, Technische Universität Berlin, Germany\\
}

}

\maketitle

\begin{abstract}
  Internet of Things (IoT) applications promise to make many aspects of our lives more efficient and adaptive through the use of distributed sensing and computing nodes.
  A central aspect of such applications is their complex communication behavior that is heavily influenced by the physical environment of the system.
  To continuously improve IoT applications, a staging environment is needed that can provide operating conditions representative of deployments in the actual production environments – similar to what is common practice in cloud application development today.
  Towards such a staging environment, we present \emph{Marvis}, a framework that orchestrates hybrid testbeds, co-simulated domain environments, and a central network simulation for testing distributed IoT applications.
  Our preliminary results include an open source prototype and a demonstration of a Vehicle-to-everything (V2X) communication scenario.
\end{abstract}


\section{Introduction}

The Internet of Things (IoT), cloud computing, and machine learning will allow for more adaptive cities, houses, and infrastructures~\cite{Jin_2014_SmartCityFramework,RisteskaStojkoska_2017_IoTforSmartHome,Mohammadi_2018_SmartCityBigML}.
However, the vision of intelligent cyber-physical systems will not be implemented with centralized cloud resources alone as they are too far away from sensor-equipped IoT devices, yielding high latencies, network bottlenecks, unnecessary energy consumption through wide-area communication, and in many use cases also considerable privacy concerns.
Addressing these limitations of central clouds, new distributed computing paradigms for the IoT such as edge and fog computing promise resources in closer proximity to sensor-equipped edge devices~\cite{Dastjerdi_2016_FogComputing}.

While continuous software testing is commonly applied in cloud environments, this is not yet the case for the emerging distributed computing environments of the IoT.
Today's cloud applications are tested through extensive use of virtualization, cluster orchestration tools, and CI/CD pipelines, allowing engineers to continuously deploy and test new software versions in so called staging environments.
These environments are set up to replicate the production environment as closely as possible to assure testing under realistic circumstances before application deployment.
However, creating such staging environments is much more challenging for IoT architectures, which are significantly more heterogeneous, distributed, and dynamically changing.
These challenges manifest themselves in a lack of adequate tools.

At the same time, continuous testing in realistic test environments is absolutely essential for many IoT applications.
For instance, if applications are to continuously optimize the operation of critical urban infrastructures such as transport systems, water infrastructures, and energy grids on the basis of collected sensor data, new versions of such applications must be tested thoroughly before they can be deployed.
It needs to be verified that an application does meet key non-functional requirements such as for its dependability and performance.
Therefore, the application behavior has to be tested under the expected distributed computing environment conditions and also variations of these conditions, given the dynamic nature of IoT environments.

In this paper, we present a first glimpse of \emph{Marvis}\footnote{Available at https://github.com/diselab/marvis}, a new framework that combines hybrid testbeds with domain-specific simulations to allow testing distributed IoT applications in adequate test environments.
Specifically, we are developing Marvis with the following requirements in mind.

\begin{enumerate}
  \item Representativity: enable testing of IoT applications in realistic conditions, so non-functional requirements such as the responsiveness and dependability can be verified.
  \item Scalability: enable testing of application behavior in IoT environments of different scales, which should not be constrained by the number of IoT devices that are available for testing.
  \item Versatility: enable testing of applications in various specific environments, so different IoT architectures and application deployments can be evaluated.
  \item Reproducibility: enable consistent testing of applications, so results can be reproduced sufficiently similar across infrastructures and multiple users.
  \item Usability: enable efficient specification, provisioning, and monitoring of the testing environments, so users can quickly test new versions of their distributed IoT applications.
\end{enumerate}

Building upon our two previous research prototypes~\cite{Boockmeyer_2019_Hatebefi,Behnke_2019_Hector}, the test environments are made up of real hardware, virtualized nodes, network simulation, and potentially multiple co-simulated domains.
All simulations run in wall clock time to facilitate test environments that behave as real as possible.
We demonstrate Marvis in this paper with a scenario, in which containers, a network simulator, and a traffic simulator are integrated automatically to test a distributed IoT application. This application connects trains and cars to level crossings on their routes via wireless communication, so that cars can adapt both their speed and potentially even routing in response to approaching trains.


\section{Related Work}\label{sec:related_work}
To test the behavior of distributed IoT applications, software engineers currently make use of various different tools such as physical testbeds~\cite{Adjih_2015_FITIoTLAB}, emulated environments~\cite{Hasenburg_2019_MockFog}, and simulators~\cite{Zeng_2017_IOTSim}.
However, these approaches usually fall short when large-scale IoT environments need to be tested with specific environment conditions, when the testing of non-functional requirements demands a certain degree of realism, or when the actual application code needs to be tested. 
Co\nobreakdash-simulation and simulation with hardware-in-the-loop (HIL) can mitigate these shortcomings by combining the initially listed tools.
Although these methods have been widely used for more than a decade in domains like automotive, power generation and distribution, or HVAC (heating, ventilation, and air conditioning)~\cite{Gomes_2018_Cosimulation_sota}, they have received little attention in the field of IoT so far.
In the following, we present and compare some existing approaches.


For verification and validation of IoT systems there exists a concept called 'Thing-In-the-Loop'~\cite{Amalfitano_2017_ThingInTheLoop}.
This model-based approach focuses on a single 'thing' under test and is therefore not comparable to Marvis, which aims to be a hybrid testbed for distributed IoT applications incorporating many devices, emulations, and simulations.
Kölsch et al. connect real world devices to the OMNET++ simulator to enable HIL simulations for IoT scenarios~\cite{Koelsch_2018_HWIL_IoT_Scenarios}.
However, in this work only a single simulation tool is addressed, whereas Marvis is a generalized framework.
A different approach, using a HIL simulator based on a multi-agent system, allows simulated and real components to join and leave the simulation at runtime to enable the co-simulation of dynamic IoT environments~\cite{Jung_2020_HWIL_Simulation_Of_IoT}.
Nevertheless, this system has limited real-time and synchronization capabilities and a distributed execution of emulators and simulators is not considered, which significantly restricts scalability.
The hybrid testbed \emph{UiTiOt}~\cite{Ly-Trong_2018_UiTiQt} follows a similar approach as Marvis, combining real IoT devices with emulated nodes, yet does not incorporate any simulators.
Furthermore, while emulations in \emph{UiTiOt} can be distributed across different virtual machines, potential latencies or bandwidth restrictions between these machines are not considered which may lead to distorted results.

In summary, to the extent of our knowledge, there exists no tool or approach that allows for the simultaneous execution of simulated, emulated, and real IoT components in a common environment which satisfies our requirements.


\section{Approach}\label{sec:approach}


Our approach in Marvis combines existing domain-specific simulators and emulators with hardware testbeds to create an environment that resembles the production environment as close as possible, thus enabling a realistic evaluation of distributed IoT applications.
Marvis creates this environment based on scenarios which specify the network, nodes, and environmental setting for the distributed applications under test.

\begin{figure}[ht]
	\centering
	\includegraphics[width=\linewidth]{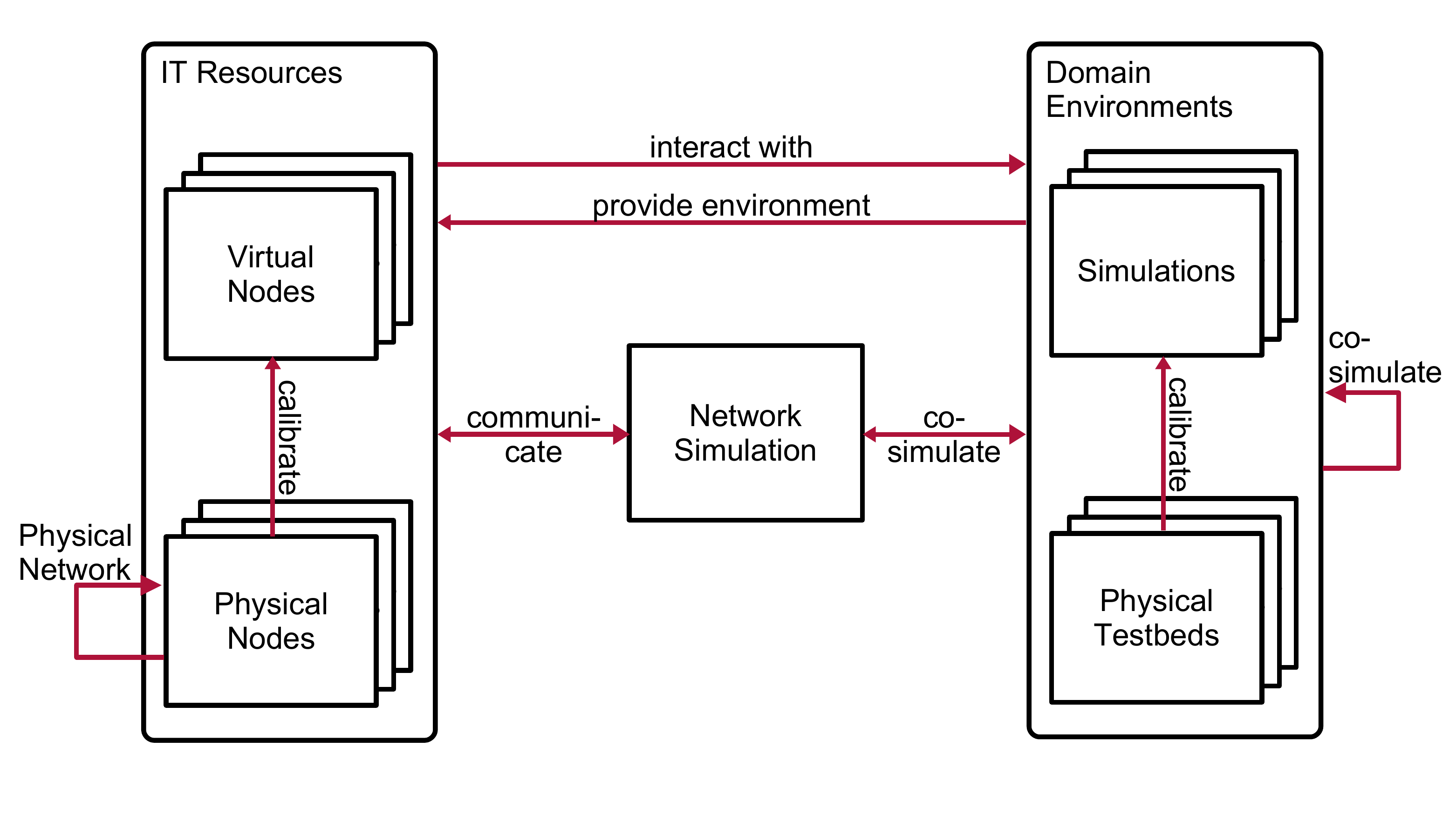}
	\caption{Marvis is a testing framework that provides realistic staging environments for distributed IoT applications. It consists of virtual and physical nodes that communicate via a virtual or physical network. The IoT applications interact with – often, but not always – simulated domain environments that are modeled after the environment in the field.}
	\label{fig:architecture}
\end{figure}

As presented in Figure~\ref{fig:architecture}, distributed IoT applications are executed on virtual or physical nodes that are connected via a network simulation or physical networks.
For the domain environments, Marvis integrates simulators, emulators, and hardware testbeds that model the environment behavior which is expected in the field.

Marvis executes all experiments in wall clock time to enable the evaluation of timing effects involving physical components.
To integrate simulators into these environments, we make use of hybrid co-simulation -- the synchronized execution of systems with discrete-event and continuous time behavior~\cite{Gomes_2018_Cosimulation_sota}.

\subsection{IT Resources}
Central to the testing framework is a network simulation which is used to model the network that is expected in the field.
The network simulation is configured with the network topology and the bandwidth and latency of the different links specified as part of the scenario description.

Marvis incorporates virtual machines and containers to provide virtual nodes on which the applications under test can be executed.
Scalability tests with the same number of nodes as in real-world IoT environments can practically only be performed with virtual nodes.
At the same time, there are effects on real hardware that are difficult to observe on virtual nodes.
The power consumption of nodes is one example where it is still difficult to simulate the effects of different application behavior.
To observe these effects, Marvis incorporates physical nodes that can be configured and connected to the network in the same way as the virtual nodes.

Virtual nodes, physical nodes, and networks can be integrated in a single scenario, allowing the testing of applications in large-scale networks while also observing their physical effect on single nodes.

\begin{figure}[ht]
	\centering
	\includegraphics[width=\linewidth]{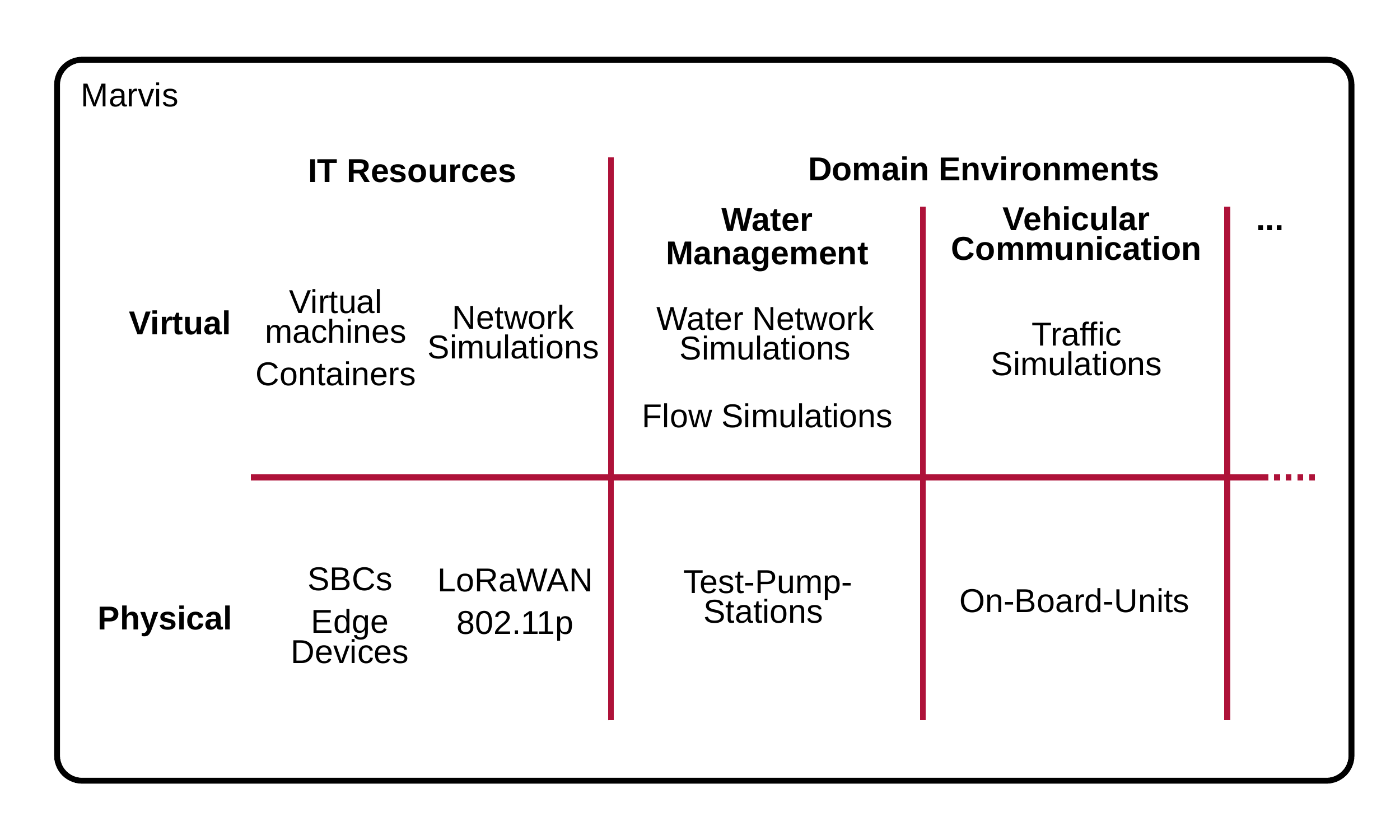}
	\caption{For each domain involved, such as water management and vehicular communication, simulators and physical testbeds constitute the environment in which applications can be executed.}
	\label{fig:components_overview_table}
\end{figure}

\subsection{Domain Environments}

Distributed IoT applications run in, and interact with, different environments which have to be modeled by our testing framework.
This interaction is seldomly one-way – which could be simulated by just feeding the output from one simulation into the next.
Rather, the integration of different domain simulators and testbeds needs to allow for mutual interactions between the different simulations and hardware.
An example for such a system is given in the evaluation, where the traffic simulator, network simulator, and the executed applications influence each other at runtime.

Marvis integrates multiple simulators and testbeds for different domain environments and facilitates the exchange of data between these simulators, testbeds, the network simulation and the nodes running the applications under test, as illustrated in Figure~\ref{fig:components_overview_table}.
While the initial Marvis design was guided by two application domains, namely water management and vehicular communication, it is extensible to different domains with their specific simulators and testbeds.

\subsection{Calibrating Virtual Systems}
Simulations of domain environments
need to be configured with many parameters to provide a realistic environment representation.
The same is true for network simulations that need to realistically represent the real-world communication behavior.
Moreover, virtualized testbed nodes modeling specific IoT devices require sensible settings with regards to available compute resources and timing behavior.

The inclusion of physical testbeds into Marvis, in form of real devices and testbeds of application domains, enables the automatic calibration of parameters by utilizing the knowledge gained from these physical components.
For this, a number of test runs on the physical
resources are leveraged to capture representative parameters for configuring virtual and simulated components.
If necessary, this process can be repeated or even run continuously.
Thus, the testing with virtual resources to enable large-scale testing can be effectively augmented by the realism of available hardware resources.

\subsection{Injecting Faults} \label{subsec:approach_fault_injection} 
In the development of hardware as well as software, fault injection testing is widely employed.
Hybrid testbeds have the potential to overcome the traditional separation between hardware and software fault injection.
We believe that this is especially beneficial for IoT applications with their software-intensive and strong cyber-physical character.


In Marvis, faults can not only be injected in the traditional fault injection points in software (e.g., modification of inputs, behavior, state) but in the environments of the applications as well.
This enables the testing of the tolerance of distributed IoT applications to external faults.
Furthermore, our framework is easily extensible to integrate software-controllable hardware components, which can then be used to inject hardware faults.
For example, a network-attached power supply can inject node failures, or a managed switch can partition the network.


\section{Preliminary Results}\label{sec:preliminary_results}

A first prototype of Marvis is used to demonstrate our approach with a specific test scenario.

\subsection{Prototype Implementation}

Marvis is predominately implemented in Python due to its library support for several simulators and networking technologies.
Consequently, test scenarios are written in Python as well to integrate the Marvis API.
Thus far, Marvis contains a co-simulation of the network simulator ns\nobreakdash-3\footnote{https://nsnam.org} and the urban mobility simulator SUMO\footnote{https://sumo.dlr.de}.
The two simulators are commonly combined for applications in the field of V2X communications~\cite{Behrisch_2011_Sumo}.
In Marvis, ns\nobreakdash-3 nodes can be associated with elements of the SUMO simulation to use, for instance, positional data to influence the network connections.

Distributed applications under test are either executed inside of docker or lxd containers as virtual nodes, on physical nodes, or a mixture of both.
Either way, all nodes are connected via ns\nobreakdash-3.
When using hardware nodes, the network interface of the simulation host is exposed to the simulation to enable the communication between the external hardware and the virtual nodes of the simulation. 


\subsection{Test Scenario}

\begin{figure}[t]
	\centering
	\includegraphics[width=0.7\columnwidth]{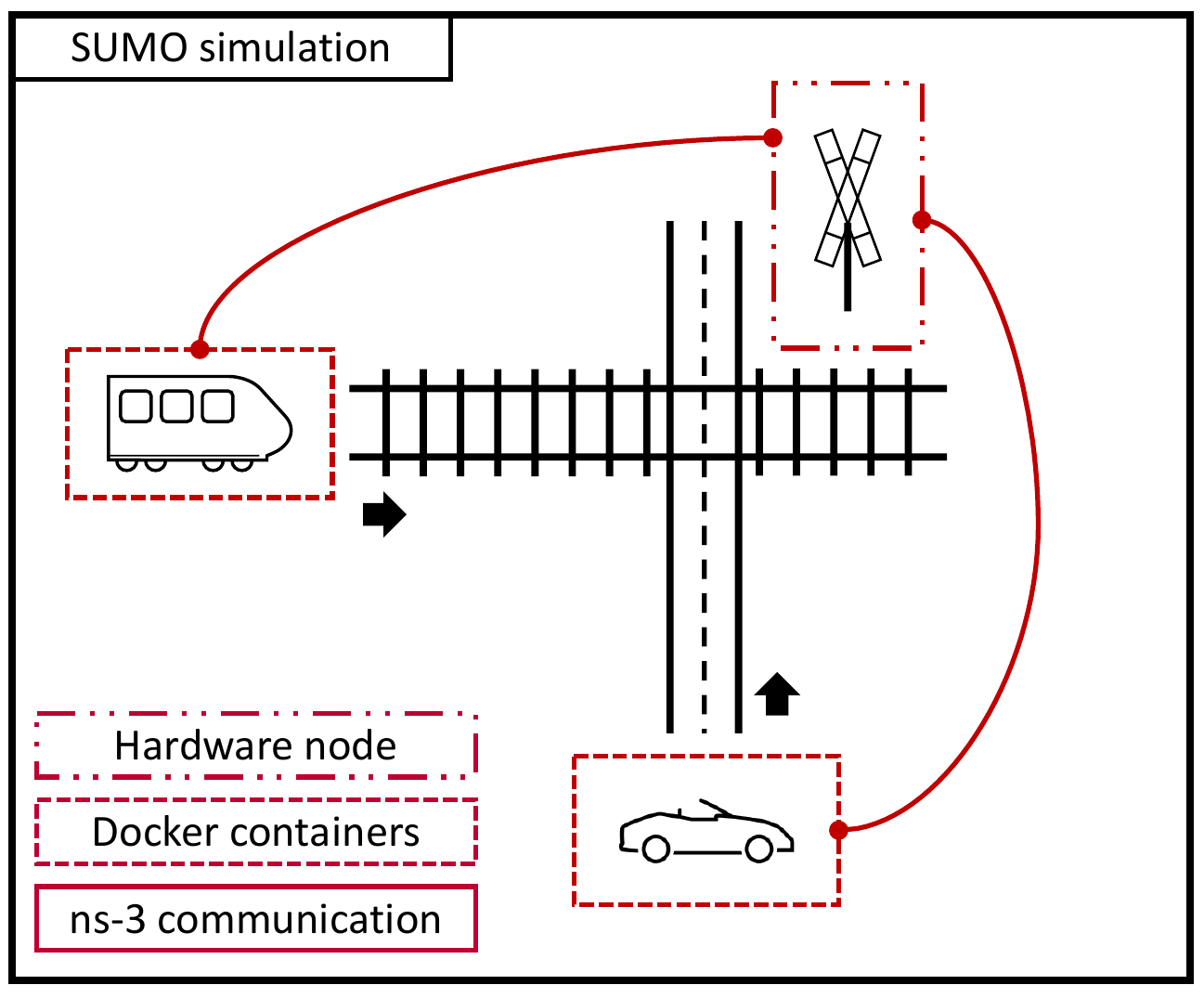}
	\caption{The train, car and level crossing, as well as the roads and trains tracks are elements of the SUMO simulation. In Marvis, the train and car are associated with docker containers, the crossing is implemented on a separate hardware node. All instances are communicating through the network simulator ns\protect\nobreakdash-3.}
	\label{fig:railcrossingscenario}
\end{figure}

We demonstrate Marvis with a test scenario that uses decentralized communication for automated driving.
As shown in Figure~\ref{fig:railcrossingscenario}, two docker containers are instantiated for the train and car, while the application of the level crossing runs on a separate hardware node.
All three elements are simulated in SUMO and connected via wireless communication in ns\nobreakdash-3.
For every simulation step in SUMO, Marvis uses the updated locations of the moving train and car to re-configure the wireless communication accordingly.
Via the V2X communication, the train can communicate its arrival to the crossing only when it is in the crossing's vicinity.
This information is then broadcast by the crossing to all cars in range.
Upon receiving this information, the car container application will stop the cars in the SUMO simulation.
The continuation of the cars' journeys is triggered after the train has passed.

When executing this scenario, the system latency, i.e. the time between the train sending a message and the car receiving it, is measured.
Naturally, this depends on the location of the train and car, which both have to be in vicinity of the crossing to be able to exchange messages.
The simulation is executed on a commodity laptop, running Ubuntu 18.04.4 LTS on an Intel Core i5-3470 with 8 GB main memory.
The applications that are run on the three elements – the train, the car and the level crossing – are executed in docker containers.
When executing this scenario, the average system latency over 100 simulation runs amounts to 10.34\,$\pm$\,1.68\,ms.
With the speed of the train of 360km/h, this means that the train is moving an average of 1.03m before the car receives the message from the approaching train.

Following this preliminary analysis, the effects of different network architectures, e.g. the demonstrated different setups with fully virtual or integrated HIL components, on the application in the domain simulation will be evaluated with Marvis.




\section{Conclusion and Future Work}\label{sec:conclusion}

This paper presented our idea and first results around Marvis, a new IoT testing framework.
Marvis integrates virtual and physical nodes and co-simulated domain environments around a central network simulation, aiming to provide a full staging environment for testing distributed IoT applications.
We implemented an early-stage research prototype of Marvis and demonstrated its usage with a realistic IoT application test scenario.

In the future, we plan to integrate more domain environments to increase the applicability of our framework.
Simultaneously, we will also evaluate our ideas with a larger set of scenarios from different application domains.
Furthermore, we are working on improving the scalability of Marvis and the methods to automatically calibrate virtual entities based on the available physical ones.

\section*{Acknowledgements}

We would like to thank Malte Andersch, Felix Gohla, Martin Michaelis, Benedikt Schenkel, and Julian Weigel for their development and research around Marvis.
Furthermore, we thank Kordian Gontarska and Daniel Richter for their comments on drafts of this paper.

\bibliographystyle{IEEEtran}
\bibliography{bibliography}

\begin{thebibliography}{10}
\providecommand{\url}[1]{#1}
\csname url@samestyle\endcsname
\providecommand{\newblock}{\relax}
\providecommand{\bibinfo}[2]{#2}
\providecommand{\BIBentrySTDinterwordspacing}{\spaceskip=0pt\relax}
\providecommand{\BIBentryALTinterwordstretchfactor}{4}
\providecommand{\BIBentryALTinterwordspacing}{\spaceskip=\fontdimen2\font plus
\BIBentryALTinterwordstretchfactor\fontdimen3\font minus
  \fontdimen4\font\relax}
\providecommand{\BIBforeignlanguage}[2]{{%
\expandafter\ifx\csname l@#1\endcsname\relax
\typeout{** WARNING: IEEEtran.bst: No hyphenation pattern has been}%
\typeout{** loaded for the language `#1'. Using the pattern for}%
\typeout{** the default language instead.}%
\else
\language=\csname l@#1\endcsname
\fi
#2}}
\providecommand{\BIBdecl}{\relax}
\BIBdecl

\bibitem{Jin_2014_SmartCityFramework}
J.~{Jin}, J.~{Gubbi}, S.~{Marusic}, and M.~{Palaniswami}, ``An information
  framework for creating a smart city through internet of things,'' \emph{IEEE
  Internet of Things Journal}, vol.~1, no.~2, 2014.

\bibitem{RisteskaStojkoska_2017_IoTforSmartHome}
B.~L. {Risteska Stojkoska} and K.~V. Trivodaliev, ``A review of internet of
  things for smart home: Challenges and solutions,'' \emph{Journal of Cleaner
  Production}, vol. 140, 2017.

\bibitem{Mohammadi_2018_SmartCityBigML}
M.~{Mohammadi} and A.~{Al-Fuqaha}, ``Enabling cognitive smart cities using big
  data and machine learning: Approaches and challenges,'' \emph{IEEE
  Communications Magazine}, vol.~56, no.~2, 2018.

\bibitem{Dastjerdi_2016_FogComputing}
A.~V. {Dastjerdi} and R.~{Buyya}, ``Fog computing: Helping the internet of
  things realize its potential,'' \emph{Computer}, vol.~49, no.~8, 2016.

\bibitem{Boockmeyer_2019_Hatebefi}
A.~{Boockmeyer}, J.~{Beilharz}, L.~{Pirl}, and A.~{Polze}, ``Hatebefi: Hybrid
  applications testbed for fault injection,'' in \emph{2019 IEEE 22nd
  International Symposium on Real-Time Distributed Computing (ISORC)}.\hskip
  1em plus 0.5em minus 0.4em\relax IEEE, 2019.

\bibitem{Behnke_2019_Hector}
I.~Behnke, L.~Thamsen, and O.~Kao, ``{H\'{e}Ctor}: A framework for testing
  {IoT} applications across heterogeneous edge and cloud testbeds,'' in
  \emph{Proceedings of the 12th IEEE/ACM International Conference on Utility
  and Cloud Computing Companion}, ser. UCC '19.\hskip 1em plus 0.5em minus
  0.4em\relax ACM, 2019.

\bibitem{Adjih_2015_FITIoTLAB}
C.~{Adjih}, E.~{Baccelli}, E.~{Fleury}, G.~{Harter}, N.~{Mitton}, T.~{Noel},
  R.~{Pissard-Gibollet}, F.~{Saint-Marcel}, G.~{Schreiner}, J.~{Vandaele}, and
  T.~{Watteyne}, ``{FIT IoT-LAB}: A large scale open experimental {IoT}
  testbed,'' in \emph{2015 IEEE 2nd World Forum on Internet of Things
  (WF-IoT)}, 2015.

\bibitem{Hasenburg_2019_MockFog}
J.~{Hasenburg}, M.~{Grambow}, E.~{Grünewald}, S.~{Huk}, and D.~{Bermbach},
  ``Mockfog: Emulating fog computing infrastructure in the cloud,'' in
  \emph{2019 IEEE International Conference on Fog Computing (ICFC)}.\hskip 1em
  plus 0.5em minus 0.4em\relax IEEE, 2019.

\bibitem{Zeng_2017_IOTSim}
X.~Zeng, S.~K. Garg, P.~Strazdins, P.~P. Jayaraman, D.~Georgakopoulos, and
  R.~Ranjan, ``{IOTSim}: A simulator for analysing {IoT} applications,''
  \emph{Journal of Systems Architecture}, vol.~72, 2017.

\bibitem{Gomes_2018_Cosimulation_sota}
C.~Gomes, C.~Thule, D.~Broman, P.~G. Larsen, and H.~Vangheluwe,
  ``Co-simulation: A survey,'' \emph{ACM Computing Survey}, vol.~51, no.~3,
  2018.

\bibitem{Amalfitano_2017_ThingInTheLoop}
D.~Amalfitano, N.~Amatucci, V.~De~Simone, V.~Riccio, and F.~A. Rita, ``Towards
  a {Thing-In-the-Loop} approach for the verification and validation of {IoT}
  systems,'' in \emph{Proceedings of the 1st ACM Workshop on the Internet of
  Safe Things}, ser. SafeThings'17.\hskip 1em plus 0.5em minus 0.4em\relax ACM,
  2017.

\bibitem{Koelsch_2018_HWIL_IoT_Scenarios}
J.~{Kölsch}, C.~{Heinz}, S.~{Schumb}, and C.~{Grimm}, ``Hardware-in-the-loop
  simulation for internet of things scenarios,'' in \emph{2018 Workshop on
  Modeling and Simulation of Cyber-Physical Energy Systems (MSCPES)}.\hskip 1em
  plus 0.5em minus 0.4em\relax IEEE, 2018.

\bibitem{Jung_2020_HWIL_Simulation_Of_IoT}
T.~Jung, N.~Jazdi, S.~Krauß, C.~Köllner, and M.~Weyrich,
  ``Hardware-in-the-loop simulation for a dynamic co-simulation of
  internet-of-things-components,'' in \emph{Proceedings of 53rd CIRP Conference
  on Manufacturing Systems 2020}, vol.~93.\hskip 1em plus 0.5em minus
  0.4em\relax Elsevier, 2020.

\bibitem{Ly-Trong_2018_UiTiQt}
N.~Ly-Trong, C.~Dang-Le-Bao, and Q.~Le-Trung, ``Towards a large-scale {IoT}
  emulation testbed based on container technology,'' in \emph{2018 IEEE Seventh
  International Conference on Communications and Electronics (ICCE)}.\hskip 1em
  plus 0.5em minus 0.4em\relax IEEE, 2018.

\bibitem{Behrisch_2011_Sumo}
M.~Behrisch, L.~Bieker, J.~Erdmann, and D.~Krajzewicz, ``{SUMO -- Simulation of
  Urban MObility}: An overview,'' in \emph{Proceedings of SIMUL 2011, The Third
  International Conference on Advances in System Simulation}.\hskip 1em plus
  0.5em minus 0.4em\relax ThinkMind, 2011.

\end{thebibliography}
\end{document}